# Optical Representation Neural Networks Enabled by Weakly Supervised Learning

*Bochang Wu[1, 2]†, Cheng-Feng Pan[1]†, Xuhao Luo[2], John You En Chan[1], Shutao Zhang[1, 2, 3], Zhenchao Liu[1], Hao Wang[4, 5], Deepshikha Arora[1], Nur Qalishah Adanan[1], Ho Soon Yee[1], Lei Chen[1], Hongtao Wang[6,7], Wang Zhang[8], Cheng-Wei Qiu[2]*, Joel K. W. Yang[1]**

*1. Engineering Product Development, Singapore University of Technology and Design, Singapore 487372, Singapore.*

*2. Department of Electrical and Computer Engineering, National University of Singapore, 4 Engineering Drive 3, Singapore 117576, Singapore.*

*3. Quantum Innovation Centre (Q.InC), Agency for Science Technology and Research (A*STAR), 2 Fusionopolis Way, Innovis #08-03, Singapore 138634, Republic of Singapore*

*4. School of Instrumentation and Optoelectronic Engineering, Beihang University, Beijing, 100191, China*

*5. Hangzhou International Innovation Institute, Beihang University, Hangzhou, 311115, China*

*6. Beijing Engineering Research Center of Mixed Reality and Advanced Display, School of Optics and Photonics, Beijing Institute of Technology, Beijing 100081, China*

*7. State Key Laboratory of Chips and Systems for Advanced Light Field Display, School of Physics, Beijing Institute of Technology, Beijing, China*

*8. College of Mechanical and Vehicle Engineering, Hunan University; Changsha 410082, P. R. China.*

**Abstract**

Diffractive optical neural networks (ONNs) show remarkable promise in high-speed and energy-efficient optical computing. However, most existing ONNs are optimized to reproduce predefined output intensity distributions, with the desired optical responses explicitly specified before training. Here, we introduce optical representation neural networks (ORNNs) enabled by coarse-grained weak supervision. Using handwritten Devanagari, Arabic, and English characters as three semantic groups, a modified contrastive objective encourages samples from the same script to cluster into similar optical representations. Remarkably, ORNNs improve the classification accuracy by approximately 9% over conventional supervised ONNs, while requiring only around 18% of the training steps to reach a stable test accuracy of 84.5%. These features can be selectively interpreted using corresponding semantic decoders. Building on this concept, we further realized semantic optical encryption with electronic neural network (ENN) decryption using experimentally acquired representations from a single-layer diffractive network fabricated by two-photon polymerization lithography (TPL). Here, we combine information from different scripts into a mixed optical ciphertext that is then first optically classified before being decrypted with a matched ENN. This work broadens the training paradigm of ONNs by coupling upstream optical representation learning with downstream task-specific decoding, opening new avenues for optical information processing.

## 1. Introduction

Diffractive optical neural networks (ONNs) have recently emerged as a promising physical computing paradigm[1-4], enabling all-optical inference with advantages in speed[5,6], parallelism[7,8], and energy efficiency[9-11] over electronic counterparts. Benefiting from these unique characteristics, optical neural networks have shown potential in image classification[12-17], holography[18-20], intelligent imaging[21-24], and optical information processing[25-28].

Despite these advances, optical neural networks are predominantly trained under supervised learning paradigms[29], where predefined optical target distributions guide the optimization process[30-33]. Such task-oriented optimization has enabled high classification accuracy and strong task-specific performance in individual optical computing tasks. [34,35]. However, because the desired optical responses are specified before training, the organization of the optical output is largely determined by externally imposed task objectives rather than natural relationships among the input samples. For example, in optical classification, each category is assigned to a predefined detector region or output location, and the optical network is optimized to reproduce this prescribed class-to-location mapping. As a result, the optical output mainly serves as the end point of a specific task, e.g. classification. In contrast, preserving the relationships among the input samples could produce intermediate representations useful for subsequent classification, interpretation, and information processing. The limitation therefore does not lie in supervised learning itself, but in the reliance of conventional task-specific ONNs on predefined optical targets, which constrain the form and spatial organization of the learned optical response. The limitation raises a fundamental question of how meaningful optical representations can form without prescribing their desired output patterns[36].

Representation learning provides a promising route for organizing high-dimensional input data into compact feature spaces that preserve meaningful relationships among samples, without explicitly prescribing the final form of the representation[37-41]. This perspective further motivates the use of weaker and more coarse-grained supervision in optical neural networks. In this work, weak supervision specifically refers to the use of script-level semantic labels during optical training, rather than individual character identities or predefined class-specific optical output targets. These coarse-grained labels provide relational information on samples that could be represented similarly, or remain distinguishable, while leaving the spatial organization of the optical representation unconstrained. Therefore, weak supervision guides the formation of the representation without explicitly prescribing its final optical form or location.

Here, we demonstrate optical representation of neural networks (ORNNs) enabled by this weakly supervised learning strategy. ORNNs are physically implemented using a diffractive optical neural network architecture. Handwritten Devanagari, Arabic, and English characters are chosen as three semantic groups with relatively similar motifs. During training, a modified contrastive objective cause script input from one language to converge into similar optical representations, while separating those from different scripts. The resulting loss is backpropagated through a differentiable optical propagation model to optimize the diffractive phase profiles with further details provided in **Supplementary Notes 1 and 2**. As a result, compact and distinguishable representations emerge for each script in the optical feature space without prescribing their output distributions or spatial locations.

The optical output also serves as an intermediate representation in addition to fulfilling its classification task. Although only coarse script-level information is used to supervise the optical network, the resulting representations retain subtle character-dependent information that can subsequently be recognized using the corresponding semantic decoders. Statistical evaluation over the complete decoder test set yields means Dice coefficients of approximately 0.73-0.77 and mean absolute Pearson correlation coefficients (|PCC|) of approximately 0.69-0.72 across the three experimental semantic categories. The framework is numerically investigated using both single-layer and three-layer diffractive architectures, while experimental validation was done only for the former. Compared with conventional supervised ONNs across multiple predefined mappings, the three-layer ORNNs achieve a mean classification accuracy of 93.4%, compared with 84.7% for supervised ONNs, while requiring approximately 82% fewer training steps to reach stable 84.5% test accuracy. Building on this capability, we further demonstrate selective semantic decoding and optical encryption, in which information associated with the three scripts is multiplexed into a scrambled ciphertext, then selectively demultiplexed, decoded and reconstructed into recognizable images. Our results show advantages in weakly supervised optical representation learning for improving training efficiency and encrypting information in security applications, where optical systems offer high-dimensional encoding and parallel information processing.[42].

## 2. Results

### 2.1 Weakly supervised optical representation learning

Figure 1a illustrates a representative task-specific supervised ONN for character recognition, while Fig. 1b introduces the proposed ORNN framework for script-level optical representation learning. In the ORNN, samples from the same script are encouraged to develop similar optical representations, while those from different scripts become distinguishable in the learned feature space.

The proposed framework for weakly supervised optical representation learning is illustrated in Fig. 1c. multi-script symbols are processed by the diffractive ONN to generate optical output fields, which are subsequently converted into compact feature representations through spatial partitioning. The calculation is then performed in the optical feature space, where samples from the same semantic category progressively organize into similar optical representations, while those from different categories become separated. This process gives rise to distinct script-dependent semantic structures without predefined optical output assignments.

The emergence of these semantic representations in the spatial optical field is conceptually illustrated in Fig. 1d. The continuous output field is represented through spatial discretization, allowing the evolving spatial organization of the optical response to be captured during learning. Rather than being assigned to predefined detector regions, semantic occupancy gradually emerges and consolidates into stable spatial patterns as optimization proceeds. Samples within the same semantic category therefore share a characteristic semantic spatial code while retaining variations in their detailed optical intensity distributions. These learned spatial codes provide the basis for the semantic interpretation and optical information processing demonstrated in the following sections.

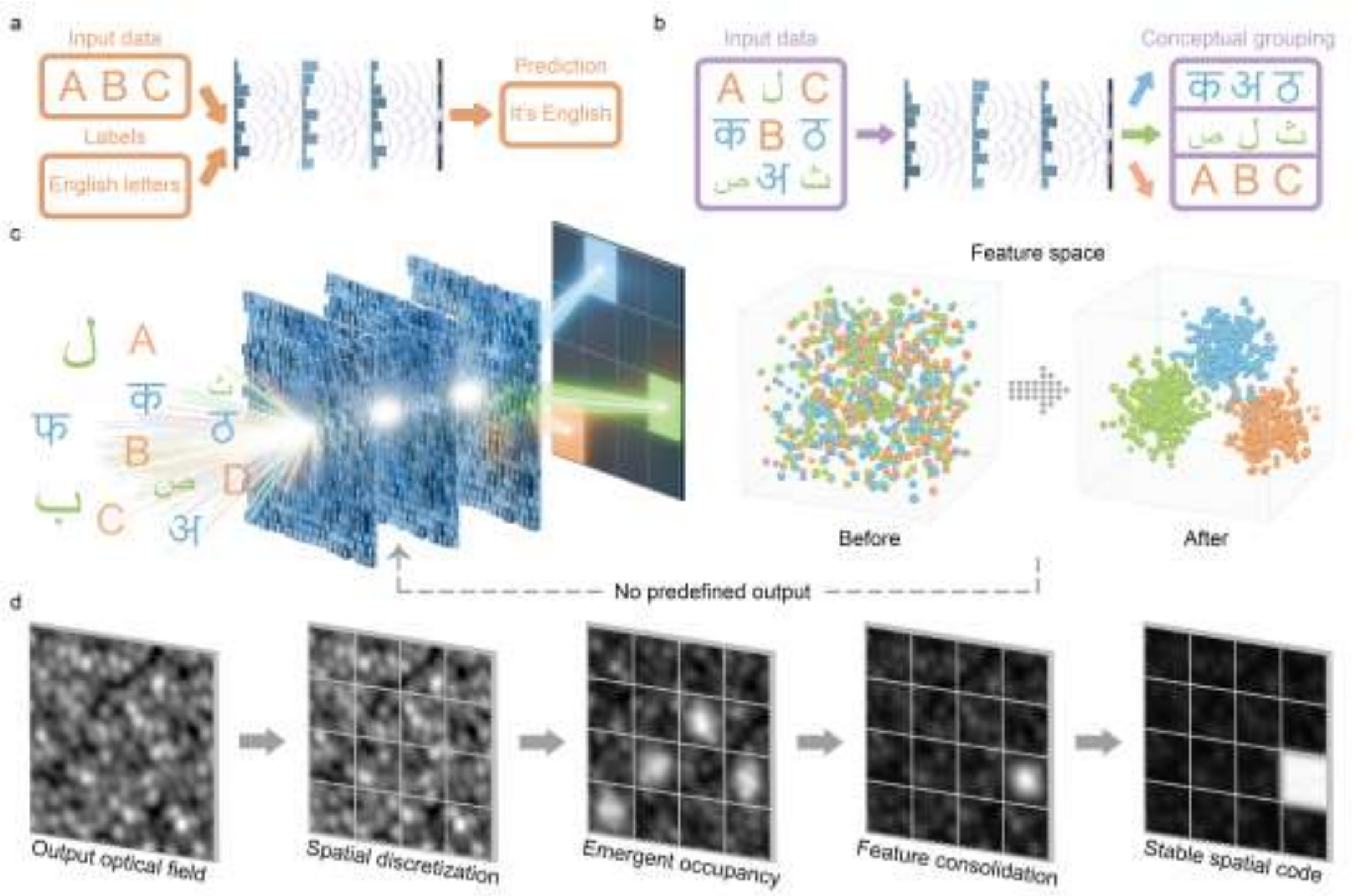

**Figure 1. Weakly supervised optical representation learning in optical representation neural networks (ORNNs). (a)** Conventional supervised ONN for character-level classification, in which labeled examples are used to train the optical network toward direct character prediction. **(b)** Conceptual illustration of the proposed ORNN for script-level optical representation learning, in which characters from different scripts are organized according to their script-level semantic relationships. **(c)** Schematic of the proposed diffractive optical framework. Input characters from Devanagari, Arabic, and English are processed by a multilayer optical network to generate spatially distributed intensity responses. Spatially extracted features form a semantic feature space in which the initially intermixed representations progressively develop distinct script-dependent organizations. **(d)** Conceptual evolution of the optical output into a stable semantic spatial code. The optical response initially occupies several candidate regions. Continued optimization progressively reinforces one dominant region, leading to feature consolidation and the formation of a stable spatial representation.

## 2.2 Formation and characterization of learned optical semantic representations

We next investigate how stable semantic clusters emerge from optical responses whose spatial distributions are not prescribed during training. The central question is whether initially unstructured optical responses will converge into consistent script-dependent spatial organization during training. To examine this process, the two-dimensional optical intensity distribution is spatially partitioned and mapped into a lower-dimensional feature representation, allowing the evolution of its spatial organization to be tracked throughout training (Fig. 2a). Driven by the modified contrastive objective, the initially dispersed feature representations progressively organize into distinct script-dependent structures during training, giving rise to learned script-dependent representations in the optical feature space.

The optical responses generated by the ONN are represented in a compact feature space and optimized using a modified contrastive learning strategy. The framework was evaluated using three publicly available handwritten character datasets, including Devanagari Handwritten Character Dataset (DHCD)[43], the Arabic Handwritten Characters Dataset (AHCD)[44,45], and the uppercase subset of the EMNIST Letters[46]. As illustrated in Fig. 2a, each optical output is spatially partitioned into an $N \times N$ grid, where $N$ is a predefined spatial-partition parameter that determines the dimensionality of the optical feature space. The average optical intensity integrated within each grid cell forms an $N^2$-dimensional feature vector, $\vec{f} = [f_1, f_2, \ldots, f_{N^2}]$, where each component represents the intensity of the corresponding spatial region. The ONN phase parameters are then optimized electronically using a modified contrastive objective defined over these feature representations. During training, the modified contrastive objective constrains only the relationships among samples, encouraging representations from the same script to become similar while separating those from different scripts. Importantly, the learned organization occurs at the script level rather than the individual-character level. Unlike conventional handwritten-character classification, where each symbol is assigned to its specific character class, our framework organizes all 36 Devanagari characters, 28 Arabic characters, and 26 English letters according to their respective scripts, forming three semantic groups.

Representations from the same semantic category are progressively drawn toward similar feature distributions, whereas those from different categories are separated. This organization is driven by the modified contrastive objective, which increases the pairwise similarity between same-script representations while reducing those between different-script representations. Importantly, this objective does not prescribe the final spatial location of any script, allowing the feature representations to self-organize into distinct script-level groups. After convergence, each script-level cluster exhibits a characteristic category-dependent feature pattern that is subsequently interpreted as a semantic spatial code. Details of the optical propagation model, network architecture, feature extraction, and learning strategy, together with the mathematical formulation of the modified contrastive objective, are provided in **Supplementary Notes 1 and 2**.

The optimized phase distributions of the three-layer diffractive network are shown in Fig. 2c. With a 5 × 5 spatial partitions, each script develops a characteristic 25-dimensional feature profile defined by the relative optical intensities across the 25 grid regions. The resulting profiles are distinct among Devanagari, Arabic, and English samples and serve as their corresponding semantic spatial codes. Although individual handwritten characters exhibit different optical interference patterns, samples from the same script share a characteristic spatial organization captured by the extracted feature vectors. Importantly, these semantic locations are not assigned before training but are identified from the learned optical responses through the relay-based semantic interpretation procedure.

To quantitatively compare the proposed learning strategy with the conventional training, we compared ORNNs with supervised ONNs in which the three script categories were randomly assigned to predefined output locations, using 9 runs per method. Multiple script-to-location mappings were evaluated for the supervised ONNs. Both methods used the same network architecture, training dataset, batch size, optimizer, and learning rate. At the shared training endpoint, the three-layer ORNNs achieved a mean test accuracy of 93.4%, compared with 84.7% for the supervised ONNs, corresponding to an improvement of approximately 8.7 % (Fig. 2d). ORNNs required 467 ± 85 training steps to reach stable 84.5% test accuracy, compared with 2534 ± 309 steps for supervised ONNs (mean ± standard deviation across nine runs per method), corresponding to approximately 82% fewer training steps (Fig. 2e). Stable attainment was defined as five consecutive evaluations at or above the threshold, with the step count taken at the first evaluation in that sequence. These results show that ORNNs achieve higher final test accuracy and reach the same accuracy threshold in fewer training steps, without prescribing script-specific output locations. Detailed results for the individual runs and the corresponding three-layer and single-layer training dynamics are provided in Supplementary Notes 3 and 4, respectively.

To determine whether the learned script-dependent organization depends on specific optical configurations, we further evaluated the framework under different spatial partition strategies and network depths. For the three-layer ORNN, numerical simulations using 4 × 4, 5 × 5, and 6 × 6 spatial partitions achieve script-level classification accuracies of approximately 94%, 95%, and 95%, respectively, while consistently producing category-dependent semantic organizations whose emergent locations vary with the feature space (Supplementary Note 5). Under the same 5 × 5 partition, the single-layer ORNN achieves approximately 91% test accuracy, approximately 3% lower than its three-layer counterpart despite the substantial decrease in optical depth (Supplementary Note 6). The same learning principle further generalizes to a distinct binary classification task using MNIST and Fashion-MNIST (Supplementary Note 7), where the single-layer ORNN achieves over 99% testing accuracy. Together, these results show that the proposed learning principle remains effective across different feature dimensions, network depths, and pattern-recognition tasks, with test accuracy exceeding 91% in the spatial-partition and network-depth evaluations and 99% on the MNIST-Fashion-MNIST task, without assigning categories to predefined output locations.

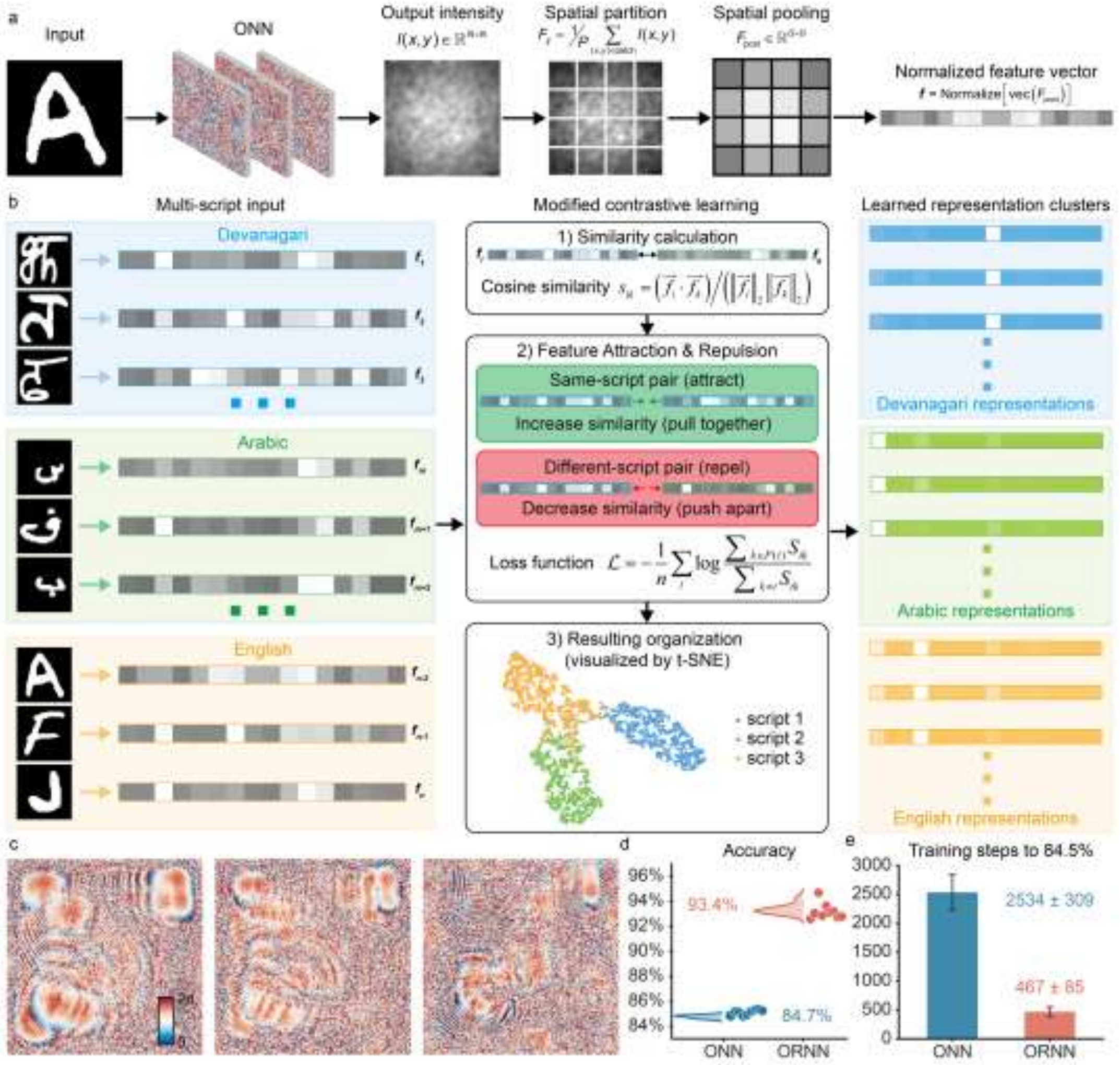

**Figure 2. Formation of optical semantic representations through weakly supervised learning. (a)** Multi-script symbols are processed by a three-layer diffractive optical neural network (ONN) to generate optical intensity distributions. The continuous output intensity distribution is spatially partitioned, and the average optical intensity within each grid cell is normalized to form a feature vector. **(b)** Training framework for weakly supervised optical representation learning. Feature vectors extracted from Devanagari, Arabic, and English samples are compared using cosine similarity and used to optimize the ONN through a modified contrastive learning strategy. Same-script representations are optimized to increase their similarity, whereas different-script representations are encouraged to decrease their similarity. Without prescribing target output locations, the optical representations progressively self-organize into distinct script-dependent feature profiles and clusters in the learned feature space. **(c)** Optimized phase distributions of the three diffractive layers obtained using a 5 × 5 grid. **(d)** Test accuracy of supervised ONNs and ORNNs at a shared training endpoint, with 9 runs per method. Individual points represent individual runs, and half-violin plots show the corresponding smooth distributions. **(e)** Training steps required to reach stable 84.5% test accuracy, defined as accuracy at or above this threshold for 5 consecutive evaluations, with the step count taken at the first evaluation in this sequence. One training step corresponds to one

optimizer update using a batch of 66 training samples. Bars and error bars indicate the mean and standard deviation across 9 runs, respectively.

## 2.3 Experimental validation of learned optical semantic representations

Having established the formation and robustness of learned optical semantic representations in simulation, we experimentally validate their preservation after physical implementation. We experimentally implemented the proposed framework using a single-layer diffractive optical neural network. The objective is to determine whether the optical semantic representations learned under weak supervision can be preserved after physical implementation and optical propagation. The experimental system consists of a digital micromirror device (DMD)-based input modulation stage, a fabricated diffractive optical element (DOE), two objective lenses, and a CMOS detector for recording output intensity distributions (Fig. 3a). Detailed experimental procedures are provided in Methods.

To implement the learned optical transformation, the optimized phase profile of the single-layer ONN for the 5 × 5 spatial partition configuration was discretized into five phase levels and fabricated as a DOE using two-photon polymerization lithography (TPL) in a Photonic GT from Nanoscribe GmbH. The fabricated DOE consisted of 200 × 200 pixels with a pixel size of 500 nm, corresponding to an overall device size of 100 × 100 μm. The fabricated diffractive phase plate and its five-level height profile are schematically illustrated in the inset of Fig. 3a, while the scanning electron microscopy (SEM) image in Fig. 3b shows the fabricated device and its surface morphology.

To evaluate the experimentally realized semantic representations, 400 output intensity distributions were captured by the CMOS detector for each of the Devanagari, Arabic, and English scripts using the 5 × 5 spatial partitions. A representative input together with its simulated and experimentally captured intensity distributions is shown in Fig. 3d-f. A video showing the experimental optical responses to input characters from all three scripts is provided in **Supplementary Video 1** and described in **Supplementary Note 23**. Although the experimental patterns contain additional background and local intensity variations compared with the simulations, the main script-dependent spatial features remain clearly visible. Interestingly, both the simulated and measured patterns frequently show enhanced intensity near the boundaries of the spatial cells, rather than forming simple isolated bright spots at the cell centers.

This boundary-enhanced behavior is likely related to the way the semantic features are defined during training. The loss operates on the average optical energy within each spatial cell and therefore does not explicitly constrain how the intensity should be distributed inside that cell. As a result, the network is free to explore different intra-cell intensity patterns as long as the pooled cell responses remain discriminative. Intensity located near a boundary can be redistributed between neighboring cells through relatively small changes in the diffractive phase profile, making these regions especially sensitive for adjusting the semantic feature vector during optimization.

The experimental preservation of the learned semantic organization was further quantified using the complete dataset of 1200 experimentally captured intensity distributions (Fig. 3c). The resulting confusion matrix yields an overall script-level classification accuracy of approximately 84%, with 400 samples from each script. The corresponding single-layer numerical model achieved approximately 91% test accuracy under the same 5 × 5 spatial partition configurations, indicating that the learned semantic organization remains well preserved in the physical system. Corresponding experimental results obtained using the 4 × 4 and 6 × 6 spatial partition configurations are provided in **Supplementary Note 9**, demonstrating that script-dependent semantic organization can also be experimentally identified under different spatial partition configurations. The 5 × 5 configuration is used for subsequent semantic-decoding experiments. These results demonstrate that learned optical semantic representations remain identifiable after fabrication and experimental optical propagation, establishing their feasibility beyond numerical simulation.

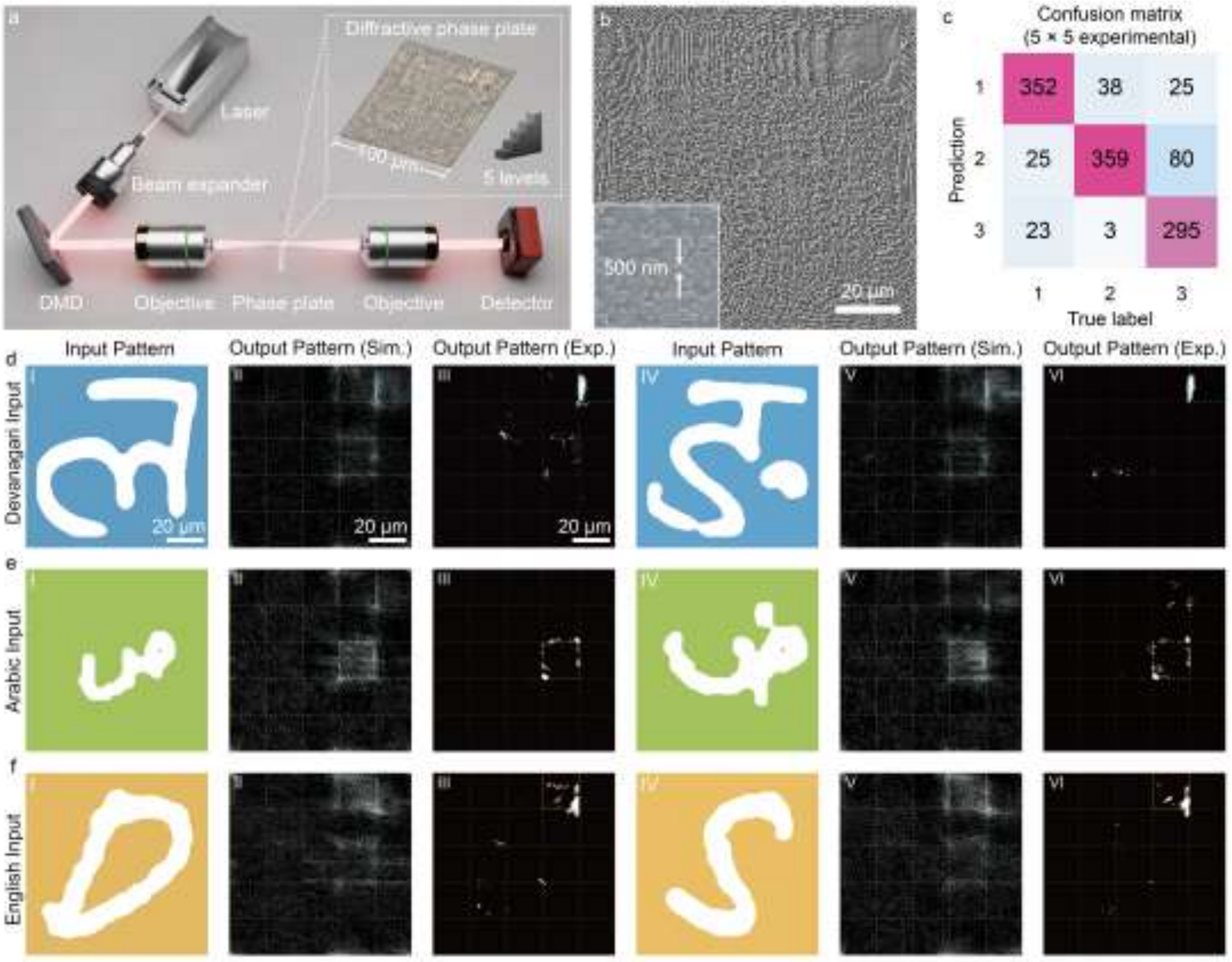

**Figure 3. Experimental demonstration of learned optical semantic representations using a single-layer diffractive optical neural network. (a)** The experimental setup comprising a laser source, beam expander, digital micromirror device (DMD), objective lenses, a fabricated diffractive phase plate, and a CMOS detector. The inset shows the diffractive phase plate and a schematic of its five-level height profile. **(b)** Scanning electron microscopy (SEM) image of the fabricated diffractive phase plate, with a magnified view of its surface features shown in the inset. **(c)** Confusion matrix obtained from 1200 experimentally captured optical intensity distributions under the 5 × 5 spatial partition configurations, with 400 samples from each script. Columns indicate the ground-truth script categories, and rows indicate the predicted categories. The overall experimental script-level classification accuracy is approximately 84%. **(d-f)** Representative input characters from **(d)** Devanagari, **(e)** Arabic, and **(f)** English, together with their corresponding simulated and experimentally captured optical intensity distributions. In each row, columns I and IV show the input characters, columns II and V show the simulated intensity distributions, and columns III and VI show the experimentally captured intensity distributions.

## 2.4 Semantic-selective decoding through semantic keys

Having established that optical semantic representations remain identifiable after physical implementation, we next examine whether these optical representations can be selectively interpreted according to their semantic identities. Classification alone demonstrates that the system can tell which script the input character belongs to but does not establish if we can determine the character from the encoded optical representation. We therefore introduce script-specific electronic neural network decoders to test whether a given optical representation can be meaningfully reconstructed only through its corresponding semantic pathway. Three independent semantic decoders were trained for the Devanagari, Arabic, and English semantic categories, serving as Key 1, Key 2, and Key 3, respectively. The decoders were trained using experimentally acquired optical semantic representations from the 5 × 5 configuration, which provided the highest experimental classification accuracy, as shown in Fig. 3. Details of the decoder architecture, reconstruction procedure, and training convergence are provided in **Supplementary Notes 10 and 11**.

Representative decoding results are shown in Fig. 4a-c, where each experimentally acquired optical semantic representation is independently processed by all three semantic keys. Across all three categories, the matched semantic key reconstructs the corresponding symbol with high structural fidelity, whereas mismatched keys produce strongly distorted outputs. Thus, the same optical representation leads to markedly different reconstruction outcomes depending on the semantic decoder that interprets it, demonstrating that meaningful reconstruction requires correspondence between the optical representation and its semantic key. The evolution of semantic reconstruction during decoder training is provided in **Supplementary Note 14**.

Semantic selectivity is further quantified using the Dice coefficient and Pearson correlation coefficient (PCC) (Fig. 4d, e). For matched semantic-key pairs, the Dice coefficients reach 0.88, 0.87, and 0.84 for the Devanagari, Arabic, and English categories, respectively, compared with 0.14-0.37 for mismatched combinations. Similarly, the corresponding PCC values reach 0.83, 0.87, and 0.79 for matched decoding, while remaining below 0.27 for mismatched keys. Consistent semantic-key selectivity is further observed using optical semantic representations obtained from simulated single-layer and three-layer ONNs, where matched decoders similarly outperform mismatched decoders in both reconstruction quality and quantitative evaluation (**Supplementary Notes 18 and 19**). The corresponding decoder training and test convergence for the experimental, single-layer simulated, and three-layer simulated representations are provided in **Supplementary Notes 11-13**, respectively. Together, these results demonstrate that learned optical semantic representations are not only distinguishable but can also be selectively interpreted through their corresponding semantic decoders, providing the basis for the semantic-level optical information processing demonstrated in the following section.

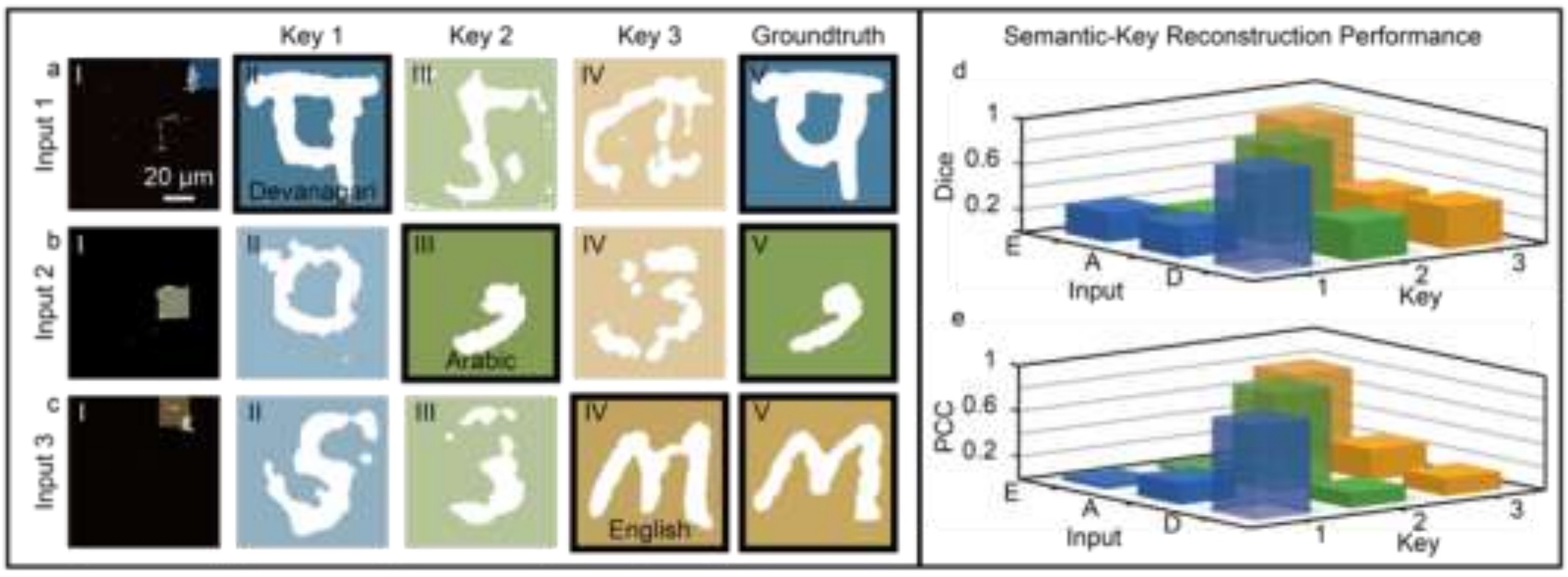

**Figure 4. Semantic-key-controlled reconstruction from optical semantic representations.** **(a-c)** Representative reconstruction results for experimentally acquired optical semantic representations belonging to the Devanagari, Arabic, and English semantic categories, respectively. In each panel, column **(I)** shows the experimentally acquired optical semantic representation generated by the diffractive optical neural network; column (**II-IV)** present the reconstruction results obtained using Key 1, Key 2, and Key 3, respectively; and column **(V)** shows the corresponding ground-truth image. The reconstruction produced by the matched semantic key is located at the diagonal position and highlighted using the same script-specific color as the corresponding ground truth. Reconstructions generated using mismatched semantic keys are displayed using desaturated colors for visual comparison. **(d,e)** Quantitative reconstruction performance evaluated using the Dice coefficient and Pearson correlation coefficient (PCC), respectively.

## 2.5 Semantic optical encryption based on learned optical representations

The semantic-key-controlled reconstruction established above further enables the learned optical representations to be used for downstream information processing. Here, we demonstrate this capability through a multi-plaintext encryption and recovery task involving three grayscale images of Newton, Maxwell, and Einstein. The three plaintexts are independently encoded using the Devanagari, Arabic, and English semantic codebooks, respectively. The resulting codewords from the three semantic channels are randomly interleaved across the ciphertext plane to form a single scrambled optical ciphertext, thereby multiplexing the three independently encoded plaintexts within a common representation (Fig. 5a).

The experimentally acquired optical responses within each script exhibit a common, self-similar spatial organization while retaining small character-specific features. These measured optical responses provide the experimentally grounded codewords used for the subsequent computational encryption and recovery process.

Following semantic routing, each channel is decoded using its corresponding semantic key, and the recovered symbols are subsequently mapped to grayscale values to reconstruct the three plain text images (Fig. 5c, d). Details of the grayscale mapping and convolutional neural network (CNN)-based plaintext reconstruction are provided in Supporting Information 21 and 22. The errors in the reconstructed plaintexts may reflect the cumulative effects of imperfect semantic reconstruction and subsequent gray-level classification. The semantic decoders first reconstruct bitmap character images from the experimentally acquired optical representations. These reconstructed images are subsequently recognized by the corresponding script-specific CNNs and mapped to one of four grayscale levels. Therefore, reconstruction errors can propagate to the subsequent gray-level classification, and an incorrect gray-level prediction directly appears as a pixel error in the final plaintext reconstruction. The corresponding final plaintext pixel accuracies, calculated by pixel-wise comparison with the original grayscale plaintexts, are approximately 72%, 68%, and 87%, respectively.

By combining experimentally acquired optical representations with semantic-key-controlled reconstruction, multiple independently encoded plaintexts can be multiplexed within a mixed ciphertext and selectively recovered through their corresponding semantic keys. This extends the learned optical representations beyond script-level organization toward semantic-aware optical information processing.

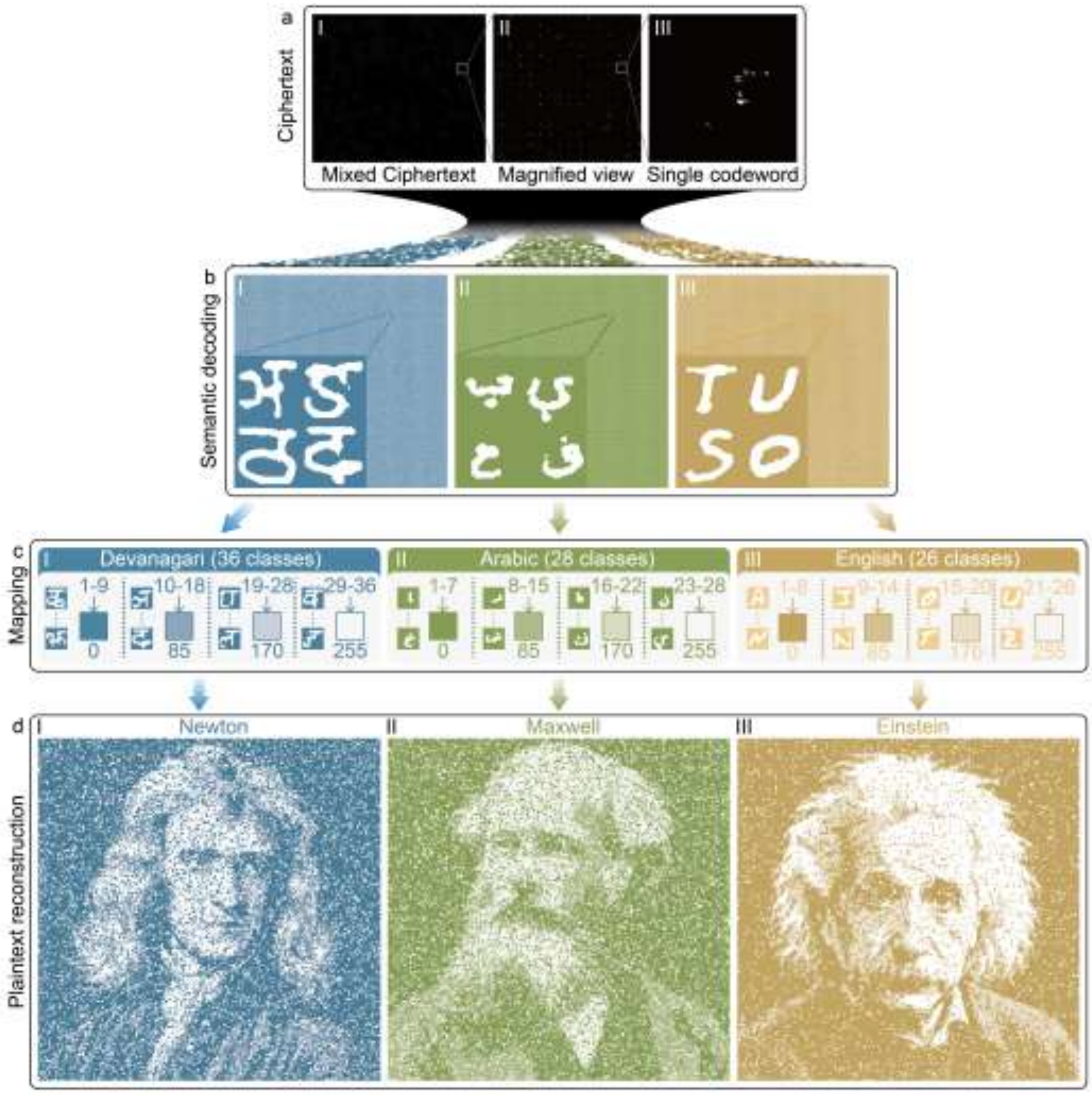

**Figure 5. Multi-semantic plaintext recovery from a mixed optical ciphertext using learned optical semantic representations. (a)** Mixed ciphertext formed by randomly interleaving codewords from three independently encoded plaintexts using the Devanagari, Arabic, and English codebooks. Enlarged views show a representative region of the mixed ciphertext and an individual optical codeword. **(b)** Computational semantic decoding of the interleaved codewords based on their experimentally acquired ONN responses. The learned script-level representations enable computational separation of the codewords into Devanagari, Arabic, and English semantic groups. **(c)** ENN-based character recognition and four-level grayscale mapping. The recovered characters are recognized by script-specific electronic neural networks, and the identified characters are mapped to grayscale values of 0, 85, 170, and 255 according to the predefined semantic codebooks. **(d)** Digital reconstruction of the Newton, Maxwell, and Einstein plaintext images by mapping the recovered symbols to their corresponding grayscale values and reassembling them at their original spatial positions.

## 3. Conclusion

In this work, we introduce optical representation neural networks (ORNNs) as an alternative learning paradigm for diffractive optical neural networks, in which the optical field is not trained to reproduce predefined script-specific targets. Instead, weak semantic supervision is used to shape the relationships between optical outputs, allowing category-dependent representations to emerge directly in the diffractive feature space. This shifts the role of the optical network from reproducing a prescribed output pattern toward learning an optical representation whose spatial organization is determined by the data and the learning objective.

Several practical limitations remain: the degradation from numerical to experimental performance likely originates from phase discretization, fabrication deviations, optical misalignment, background scattering, and detector noise. Network depth also introduces a trade-off: a single diffractive layer simplifies fabrication and alignment but provides fewer degrees of freedom for optical field transformation than a three-layer architecture, resulting in reduced representation capacity and slower convergence. In addition, the semantic-encryption demonstration still exhibits imperfect reconstruction arising from variations in the learned optical representations and downstream decoding errors and should therefore be regarded as a proof-of-concept demonstration of semantic-level optical information processing.

An important consequence of this formulation is that the learned optical field is not tied to a single downstream decision. The same representation can be interpreted, decoded, routed, or reused for subsequent information-processing operations, while its spatial organization is learned under weak supervision without predefined output locations. The persistence of this behavior across different spatial partitions and network depths, together with its experimental realization in a fabricated diffractive system, suggests that the observed organization is not simply a consequence of a particular output geometry or numerical configuration.

The semantic-key-controlled reconstruction and optical information-processing demonstrations further illustrate how such learned representations can serve as functional optical states rather than task-specific classification outputs. More broadly, these results suggest a route toward diffractive optical systems in which representation formation and downstream processing are separated: the optical front end first learns a physically encoded semantic space, while subsequent operations can be designed around the learned representation rather than requiring the optical network itself to be retrained for every prescribed output pattern. The present framework still relies on weak category-level supervision and computational interpretation of the learned representations. Extending this concept toward weaker supervision, larger and more continuous semantic spaces, and increasingly optical implementations of representation interpretation and downstream processing may provide further opportunities for developing adaptive optical information-processing systems in which useful physical representations emerge from data rather than being specified in advance.

**4. Experimental Section. Experimental Section**

**Supporting Information**

Supporting Information is available from the Wiley Online Library or from the author.

**Acknowledgements**

**Conflict of interest**

The authors declare that they have no competing interests.

**Author contributions**

B.C.W., C.-F.P., C.-W.Q., and J.K.W.Y. conceived the idea. B.C.W. and X.H.L developed the methodology. B.C.W. performed simulations and experiments. B.C.W. and C.-F.P. carried out the device fabrication. B.C.W., C.-F.P., Z.C.L., and H.W. performed the characterization and data analysis. B.C.W., S.T.Z. and C.-F.P. prepared schematic illustrations. C.-W.Q. and J.K.W.Y. supervised the project. S.T.Z., X.H.L., D.A., N.Q.A., H.S.Y., L.C., H.T.W., W.Z., C.-W.Q., and J.K.W.Y. provided valuable discussions and suggestions on the manuscript. B.C.W. wrote the manuscript, and all authors reviewed and revised the manuscript.

**Data availability**

The data that support the findings of this study are available from the corresponding author upon reasonable request.

## Reference

1. Lin, X. *et al.* All-optical machine learning using diffractive deep neural networks. *Science* **361**, 1004-1008 (2018).
2. Liu, C. *et al.* A programmable diffractive deep neural network based on a digital-coding metasurface array. *Nat. Electron.* **5**, 113-122 (2022).
3. Rahman, M. S. S., Li, Y., Yang, X., Chen, S. & Ozcan, A. Massively parallel and universal approximation of nonlinear functions using diffractive processors. *eLight* **5**, 32 (2025).
4. Wang, Y. *et al.* Reconfigurable versatile integrated photonic computing chip. *eLight* **5**, 20 (2025).
5. Shen, Y. *et al.* Deep learning with coherent nanophotonic circuits. *Nat. Photonics* **11**, 441-446 (2017).
6. Xu, X. *et al.* 11 TOPS photonic convolutional accelerator for optical neural networks. *Nature* **589**, 44-51 (2021).
7. Feldmann, J. *et al.* Parallel convolutional processing using an integrated photonic tensor core. *Nature* **589**, 52-58 (2021).
8. Fu, T. *et al.* Photonic machine learning with on-chip diffractive optics. *Nat. Commun.* **14**, 70 (2023).
9. Fu, T. *et al.* Optical neural networks: progress and challenges. *Light Sci. Appl.* **13**, 263 (2024).
10. Hughes, T. W., Minkov, M., Shi, Y. & Fan, S. Training of photonic neural networks through in situ backpropagation and gradient measurement. *Optica* **5**, 864 (2018).
11. Dong, Y., Bai, Y., Zhang, Q., Luan, H. & Gu, M. High-throughput optical neuromorphic graphic processing at millions of images per second. *eLight* **5**, 29 (2025).
12. Chang, J., Sitzmann, V., Dun, X., Heidrich, W. & Wetzstein, G. Hybrid optical-electronic convolutional neural networks with optimized diffractive optics for image classification. *Sci. Rep.* **8**, 12324 (2018).
13. Ashtiani, F., Geers, A. J. & Aflatouni, F. An on-chip photonic deep neural network for image classification. *Nature* **606**, 501-506 (2022).
14. Cheng, J. *et al.* Multimodal deep learning using on-chip diffractive optics with in situ training capability. *Nat. Commun.* **15**, 6189 (2024).
15. Mengu, D., Luo, Y., Rivenson, Y. & Ozcan, A. Analysis of Diffractive Optical Neural Networks and Their Integration with Electronic Neural Networks. *IEEE J. Sel. Top. Quantum Electron. Publ. IEEE Lasers Electro-Opt. Soc.* **26**, 3700114 (2020).
16. Hu, J. *et al.* Diffractive optical computing in free space. *Nat. Commun.* **15**, 1525 (2024).
17. Huang, Z. *et al.* Anti-interference diffractive deep neural networks for multi-object recognition. *Light Sci. Appl.* **15**, 101 (2026).
18. Sakib Rahman, M. S. & Ozcan, A. Computer-Free, All-Optical Reconstruction of Holograms Using Diffractive Networks. *ACS Photonics* **8**, 3375-3384 (2021).
19. Hu, J. *et al.* Ultrathin, Wavelength-Multiplexed and Integrated Holograms and Optical Neural Networks Based on 2D Perovskite Nanofilms. *Laser Photonics Rev.* **19**, 2401458 (2025).
20. Fan, Z. *et al.* Holographic multiplexing metasurface with twisted diffractive neural network. *Nat. Commun.* **15**, 9416 (2024).
21. Işıl, Ç. *et al.* Super-resolution image display using diffractive decoders. *Sci. Adv.* **8**,

eadd3433 (2022).
22. Chen, S., Li, Y., Wang, Y., Chen, H. & Ozcan, A. Optical generative models. *Nature* **644**, 903-911 (2025).
23. Işıl, Ç. *et al.* All-optical image denoising using a diffractive visual processor. *Light Sci. Appl.* **13**, 43 (2024).
24. Hu, J. *et al.* Subwavelength imaging using a solid-immersion diffractive optical processor. *eLight* **4**, 8 (2024).
25. He, C. *et al.* Orbital angular momentum multiplexing diffractive neural networks for high-capacity optical inference. *Nat. Photonics* 1-9 (2026) doi:10.1038/s41566-026-01930-2.
26. Li, J. *et al.* All-optical complex field imaging using diffractive processors. *Light Sci. Appl.* **13**, 120 (2024).
27. Yu, H. *et al.* All-optical image transportation through a multimode fibre using a miniaturized diffractive neural network on the distal facet. *Nat. Photonics* **19**, 486-493 (2025).
28. Yu, X. *et al.* Parallel optical computing capable of 100-wavelength multiplexing. *eLight* **5**, 10 (2025).
29. Chen, H. *et al.* Diffractive deep neural networks: Theories, optimization, and applications. *Appl. Phys. Rev.* **11**, 021332 (2024).
30. Sui, X., Wu, Q., Liu, J., Chen, Q. & Gu, G. A Review of Optical Neural Networks. *IEEE Access* **8**, 70773-70783 (2020).
31. Luo, Y. *et al.* Design of task-specific optical systems using broadband diffractive neural networks. *Light Sci. Appl.* **8**, 112 (2019).
32. Duan, Z., Chen, H. & Lin, X. Optical multi-task learning using multi-wavelength diffractive deep neural networks. *Nanophotonics* **12**, 893-903 (2023).
33. Liu, W. *et al.* Photonic Mixture-of-Experts for scalable multi-task on-chip optical neural networks. *Nat. Commun.* https://doi.org/10.1038/s41467-026-73983-4 (2026) doi:10.1038/s41467-026-73983-4.
34. Zhang, H. *et al.* Integrated platforms and techniques for photonic neural networks. *Npj Nanophotonics* **2**, 40 (2025).
35. Wetzstein, G. *et al.* Inference in artificial intelligence with deep optics and photonics. *Nature* **588**, 39-47 (2020).
36. Lin, Z. *et al.* End-to-end metasurface inverse design for single-shot multi-channel imaging. *Opt. Express* **30**, 28358-28370 (2022).
37. Radford, A. *et al.* Learning Transferable Visual Models From Natural Language Supervision. in *Proceedings of the 38th International Conference on Machine Learning* 8748-8763 (PMLR, 2021).
38. Pan, S. J. & Yang, Q. A Survey on Transfer Learning. *IEEE Trans. Knowl. Data Eng.* **22**, 1345-1359 (2010).
39. Yosinski, J., Clune, J., Bengio, Y. & Lipson, H. How transferable are features in deep neural networks? in *Advances in Neural Information Processing Systems* vol. 27 (Curran Associates, Inc., 2014).
40. Bengio, Y., Courville, A. & Vincent, P. Representation Learning: A Review and New Perspectives. *IEEE Trans. Pattern Anal. Mach. Intell.* **35**, 1798-1828 (2013).
41. LeCun, Y., Bengio, Y. & Hinton, G. Deep learning. *Nature* **521**, 436-444 (2015).

42. Wang, H. *et al.* Optical information security. *Nat. Rev. Phys.* 1-17 (2026) doi:10.1038/s42254-026-00980-3.
43. Acharya, S., Pant, A. K. & Gyawali, P. K. Deep learning based large scale handwritten Devanagari character recognition. in *2015 9th International Conference on Software, Knowledge, Information Management and Applications (SKIMA)* 1-6 (2015). doi:10.1109/SKIMA.2015.7400041.
44. El-Sawy, A., EL-Bakry, H. & Loey, M. CNN for Handwritten Arabic Digits Recognition Based on LeNet-5. in *Proceedings of the International Conference on Advanced Intelligent Systems and Informatics 2016* (eds Hassanien, A. E., Shaalan, K., Gaber, T., Azar, A. T. & Tolba, M. F.) 566-575 (Springer International Publishing, Cham, 2017). doi:10.1007/978-3-319-48308-5_54.
45. AlJarrah, M. N., Zyout, M. M. & Duwairi, R. Arabic Handwritten Characters Recognition Using Convolutional Neural Network. in *2021 12th International Conference on Information and Communication Systems (ICICS)* 182-188 (2021). doi:10.1109/ICICS52457.2021.9464596.
46. Cohen, G., Afshar, S., Tapson, J. & van Schaik, A. EMNIST: Extending MNIST to handwritten letters. in *2017 International Joint Conference on Neural Networks (IJCNN)* 2921-2926 (2017). doi:10.1109/IJCNN.2017.7966217.